\documentclass[]{aastex631}

\usepackage{natbib}
\usepackage{array}
\usepackage{graphicx}
\usepackage{latexsym}
\usepackage{amsfonts,amsmath,amssymb}
\usepackage{url}
\usepackage[utf8]{inputenc}
\usepackage{hyperref}
\usepackage{placeins}
\hypersetup{colorlinks=true,citecolor=blue,linkcolor=blue,urlcolor=blue,breaklinks=true}
\usepackage[shortlabels]{enumitem} %

\shorttitle{Quantitative Criteria for Defining Planets}
\shortauthors{Margot et al.}

\turnoffedit

\begin{document}

\title{Quantitative Criteria for Defining Planets}

\correspondingauthor{Jean-Luc Margot}
\email{jlm@epss.ucla.edu}

\author[0000-0001-9798-1797]{Jean-Luc Margot}
\affiliation{Department of Earth, Planetary, and Space Sciences, University of California, Los Angeles, CA 90095, USA}
\affiliation{Department of Physics and Astronomy, University of California, Los Angeles, CA 90095, USA}

\author[0000-0002-0283-2260]{Brett Gladman}
\affiliation{Department of Physics and Astronomy, University of British Columbia, Vancouver, V6T 1Z1, Canada}

\author{Tony Yang}
\affiliation{Chaparral High School, Temecula, CA 92591, USA}

\begin{abstract}
  {The current IAU definition of ``planet'' is
problematic because it is vague and excludes exoplanets.  Here, we
describe aspects of quantitative planetary taxonomy and examine the
results of unsupervised clustering of Solar System bodies to guide
the development of possible classification frameworks.  Two
unsurprising conclusions emerged from the clustering analysis: (1)
satellites are distinct from planets and (2) dynamical dominance is
a natural organizing principle for planetary taxonomy.  To
generalize an existing dynamical dominance criterion, we adopt a
universal clearing timescale applicable to all central bodies (brown
dwarfs, stars, and stellar remnants).  Then, we propose two
quantitative, unified frameworks to define both planets and
exoplanets.  The first framework is aligned with both the IAU
definition of planet in the Solar System and the IAU working
definition of an exoplanet.  The second framework is a simpler
mass-based framework that avoids some of the difficulties ingrained
in current IAU recommendations.}
\end{abstract}
\bibliographystyle{psj}

\section{Introduction}
\label{sec-intro}
In 2006, the International Astronomical Union (IAU) adopted resolution
B5, which contains the following definition:
\begin{quote}
A planet is a celestial body that (a) is in orbit
around the Sun, (b) has sufficient mass for its self-gravity to
overcome rigid body forces so that it assumes a hydrostatic
equilibrium (nearly round) shape, and (c) has cleared the
neighbourhood around its orbit.
\end{quote}
The current IAU definition of ``planet'' is problematic both because
it is not quantitative and because it excludes exoplanets.
In a previous paper, one of
us proposed a possible solution to remedy both problems
\citep{marg15planet}.

In 2018, IAU Commission F2 “Exoplanets and the Solar System”
promulgated the following working definition for exoplanets:
\begin{quote}
  Objects with true masses below the limiting mass for thermonuclear
  fusion of deuterium (currently calculated to be 13 Jupiter masses
  for objects of solar metallicity) that orbit stars, brown dwarfs or
  stellar remnants and that have a mass ratio with the central object
  below the L4 / L5 instability ($M/M_{\rm central} < 2/(25 + \sqrt{621})
  \simeq 1/25$) are “planets”, no matter how they formed.  The minimum
  mass/size required for an extrasolar object to be considered a
  planet should be the same as that used in our Solar System, which is
  a mass sufficient both for self-gravity to overcome rigid body
  forces and for clearing the neighborhood around the object’s orbit.
\end{quote}
The rationale for this working definition is explained in detail by
\citet{leca22}.  Notably, the minimum mass/size required for an
extrasolar object to be considered a planet remains undefined because
IAU resolution B5 does not define it {precisely, except through
  vague implications about the mass required to clear a zone (how
  large? how clear?) and to approach hydrostatic equilibrium
  (how closely?).}

{In this paper, we present two perspectives to stimulate
  conversations about improving the current IAU planet definition.
  Both proposals include a quantitative, unified framework to define
  both planets and exoplanets.  The first proposal is aligned with the
  spirit of IAU Resolution B5 and incorporates the working definition
  of IAU Commission F2.  The second proposal is a simpler mass-based
  criterion.  In both instances, we follow a common approach to
  astronomical nomenclature, which is to use physical principles to
  establish meaningful thresholds and then adhere to the thresholds
  for definitional purposes.}

We outline a metric for
dynamical dominance in Section~\ref{sec-Pi}, describe desirable
features of a planetary taxonomy in Section~\ref{sec-practical}, use
unsupervised clustering techniques to guide a proposed planetary
taxonomy in Section~\ref{sec-clustering}, and extend an existing
dynamical dominance criterion in Section~\ref{sec-simpler}.  {We
propose an IAU-aligned taxonomic framework in
Section~\ref{sec-proposal} and a simpler mass-based framework in
Section~\ref{sec-simplified}.}

\section{Dynamical Dominance}
\label{sec-Pi}

{Dynamical dominance looms large in planetary taxonomy because
  both Ceres and Pluto lost their status as planets once they were
  found to belong to a belt of small bodies}.
\citet{marg15planet} developed the planetary discriminant $\Pi$ to
quantify what it takes to clear a well-defined orbital zone around a
planetary body, echoing the IAU definition.  A logical choice for the
extent of the zone to be cleared around a planetary body is the
canonical feeding zone~\citep{birn73,arty87,ida93,glad93}, which is
$2\sqrt{3}$ times the Hill radius of the planetary body:
\begin{equation}
r_{H} = \left( \frac{m}{3m_{\rm central}} \right)^{1/3}a,
\label{eq-rh}
\end{equation}
where $m$ is the mass of the planetary body, $m_{\rm central}$ is the
mass of the central body or host star, and $a$ is the semimajor axis.

Following \citet{trem93}, one can model the ejection of planetesimals
by a dynamically dominant body as a diffusion process and derive the
minimum orbit-clearing mass required to clear a zone of extent $C$
times the Hill radius in a
given clearing timescale
$t_{\rm clear}$ \citep[][Equation 8]{marg15planet}:
\begin{equation}
  m _{\rm clear} = C^{3/2} m_{\rm central}^{5/8} \left(\frac{t_{\rm clear}}{ 1.1\times10^5\, {\rm y}}\right)^{-3/4} a^{9/8}, 
\label{eq-mclear}
\end{equation}
where {$a$ is expressed in au and} $m _{\rm clear}$, $m_{\rm central}$, and $t_{\rm clear}$ are
expressed in units of earth masses, solar masses, and years,
respectively.  {This expression was derived for definitional
  purposes in the context of circular orbits and is not meant to
  capture all astrophysical situations. Clearing in the context of
  eccentric orbits does not yield a simple expression
  \citep[e.g.,][]{quil06,morr15}. Likewise, additional perturbations
  due to other planets, dynamical friction between the smaller
  bodies, and the effects of gas drag and disk tides are not
  included.}

The parameter $\Pi$ is simply
the mass of a planetary body expressed
in units of the
corresponding orbit-clearing mass:
\begin{equation}
\Pi = \frac{m}{m_{\rm clear}}.
\end{equation}
Values of $\Pi$ larger than 1 convey an ability to clear the feeding
zone, whereas values of $\Pi$ smaller than 1 convey an inability to do
so.  In the Solar System with a clearing timescale $t_{\rm clear}=$10
billion years and $C=2\sqrt{3}$, the expressions for the minimum orbit-clearing mass
and $\Pi$ reduce to
\begin{equation}
m _{\rm clear} = 0.001239\, a^{9/8}, 
\label{eq-mclearss}
\end{equation}
and 
\begin{equation}
\Pi = 807\, ma^{-9/8},
\label{eq-Piss}
\end{equation}
where $m$ is the mass of the planetary body in Earth masses and $a$ is
the semimajor axis in au.

{We emphasize the focus on {\em the ability to clear a zone} in
  a specified timescale as opposed to {\em the state of having cleared
    a zone}.  This distinction is important because dynamically
  dominant bodies retain their ability to clear a zone -- and their
  status as planets -- even if small bodies enter the zone or if the
  zone is still in the process of being cleared.  In other words, the
  ability to clear a zone is relatively impervious to various
  evolutionary phases in the lives of planets (e.g., early formation
  epoch, Nice-like migration, gravitational scattering).}

\section{Practical considerations for a taxonomy of planetary bodies}
\label{sec-practical}
We would like our taxonomy of planetary bodies to be {\em useful} and
{\em consistent}.  A taxonomy is useful if it helps guide, or
contributes to, our scientific understanding of the population.
Conversely, a taxonomy that obscures fundamental relationships among
specimens is not useful.  In order to be useful, a taxonomy must also
be based on features that are readily observable and measurable.  A taxonomy
based on unobservable or difficult-to-observe properties
has no value.

Shortly after the discovery of a planetary system, the first
properties that we can measure are the orbital elements, starting with
the orbital periods and semimajor axes.  In many situations, we can
then obtain mass or minimum-mass estimates, either from radial
velocity measurements, transit-timing variations, orbital
perturbations, or the presence of satellites.  In some situations,
size is easier to measure than mass, either from transit depths,
secondary eclipses, or optical/IR flux measurements.  In these
instances, a radius--mass relationship can be used to provide a mass
estimate.  Popular radius--mass relationships and their domain of
applicability include ~\citet[][$R_p<25R_\earth$]{fang12exostats},
\citet[][$R_p<11R_\earth$]{wu13}, \citet[][$R_p<4R_\earth$]{weis14},
\citet{fabr14}, \citet[][$R_p<4R_\earth$]{wolf15}, \citet[][9 orders
  of magnitude]{chen17}, \citet[][$M_p<120M_\earth$]{oteg20}.  To
summarize, we can generally expect measurements of orbital properties
and masses shortly after discovery.

In contrast, the shape of planetary bodies in newly discovered
planetary systems cannot be determined with technology available now
or in the foreseeable future.  Lightcurve information can sometimes
provide useful limits on an object's axial ratio and convex shape.
However, for distant objects where the viewing and illumination
geometries change slowly, the spin orientation and 3D shape remain
undetermined.  For instance, the shape of the $\sim$360 km diameter
Neptunian moon Nereid remains undetermined (``somewhere between
roughly spherical and greatly nonspherical''), despite an extensive
set of $\sim$600 lightcurves spanning 20 years \citep{scha08}.
Stellar occultations that involve multiple chords can provide a 2D
projection of an object's shape at the occultation epoch.  Although
quite useful, these data are usually insufficient to make a confident
determination about the 3D shape of an object~\citep{ort20}.  Because
shapes are so difficult to measure,
even the shapes of the largest trans-Neptunian bodies at mere
distances of $\sim$40 au remain enigmatic.
For
instance, the shapes of the $\sim$920~km diameter Orcus and
$\sim$850~km diameter Salacia remain uncertain, despite discoveries
$\sim$20 years ago \citep{grun19,emer23}.  For exoworlds, the problem
is much worse because the spatial resolution is degraded by a factor
of $\sim$10,000 compared to the trans-Neptunian region, even for the
nearest stars.
Therefore, shape information is
either difficult or impossible to obtain,
so planetary taxonomies that rely on the knowledge of shapes are
unworkable.

\begin{figure}[h]
\begin{center}
\includegraphics[width=\columnwidth]{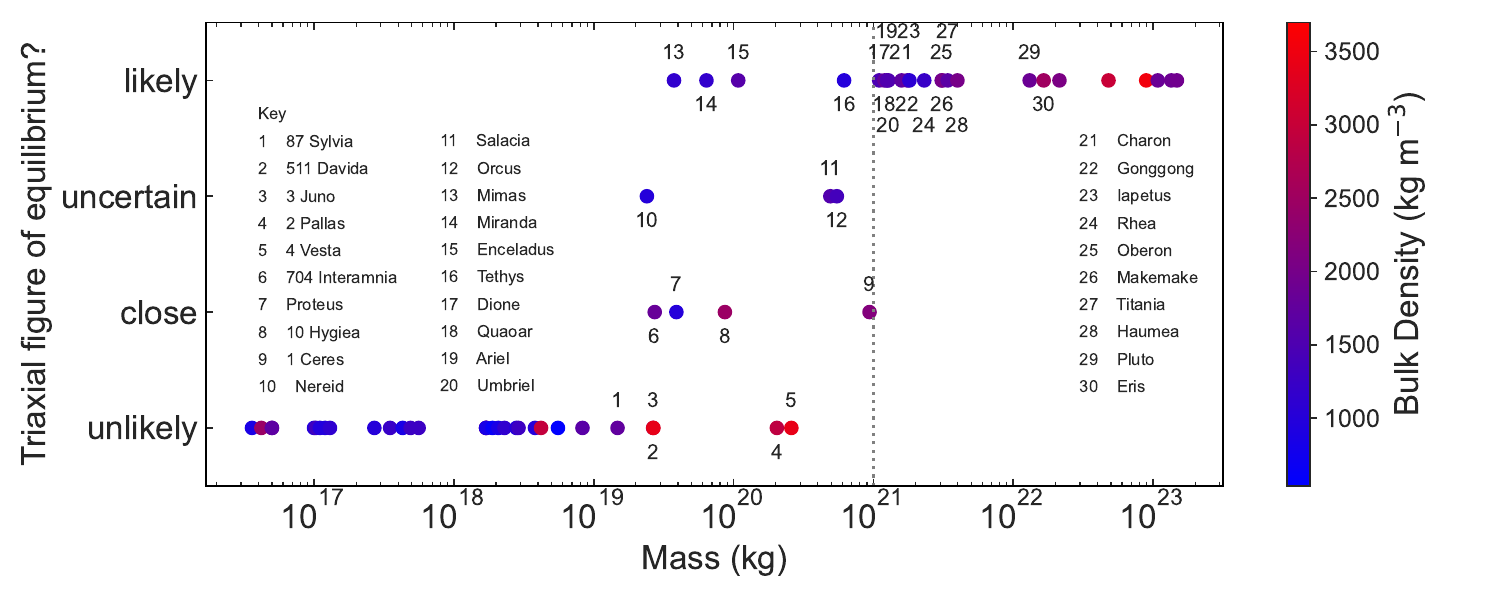}
\caption{%
  Likelihood of adopting a triaxial figure of equilibrium for 62 Solar
  System bodies as a function of mass.  Objects listed as ``close''
  have been recognized as approaching but clearly deviating from an
  hydrostatic equilibrium shape.  Objects in the
  $10^{19}-2\times10^{22}$ kg range are labeled.  The vertical dotted
  line represents a possible mass threshold at 10$^{21}$ kg.}
\label{fig-round}
\end{center}
\end{figure}

Taxonomic classification may be aided by the presence of gaps between
individual clusters of planetary bodies.  As described by
\citet{sote06}, nature does not provide an obvious gap between
spheroidal and nonspheroidal shapes (Figure~\ref{fig-round}).  The
transition between these two regimes depends on multiple properties,
including bulk density and material strength \citep{tanc08}, as well
as collisional, tidal, and thermal evolution histories.  Most of these
properties are not observable remotely.  Transitional objects exhibit
a range of mass values spanning $\sim$2 orders of magnitude, such that
mass (or diameter) is not a good proxy for roundness.  However, it
appears that Solar System objects with masses larger than $10^{21}$ kg
have sufficient mass to be approximately in hydrostatic equilibrium
and adopt a nearly triaxial figure of equilibrium
(Figure~\ref{fig-round}).
This mass threshold, which corresponds approximately to the masses of
Ceres and Dione, could perhaps be used if one had to make
{an informed guess} about the state of hydrostatic equilibrium of a
planetary body in the absence of detailed shape information.

\section{Unsupervised clustering as a guide to taxonomic classification}
\label{sec-clustering}
Taxonomists usually group specimens according to certain features.
However, there are multiple ways to delineate groups in a taxonomy of
planetary systems, and we would like to establish the groups in a way
that is both logical and dispassionate.  Unsupervised clustering
algorithms are
{useful} in this context, as they
can guide our choices impartially on the basis of well-established
algorithms {\citep{kauf90,ever11}}.
Because no training is involved in unsupervised algorithms, there is
no mechanism for intentionally or unintentionally biasing the results.

\subsection{K-means clustering}
K-means clustering is a popular {nonparametric} algorithm designed to partition $n$
observations ($\boldsymbol{x_1}, ..., \boldsymbol{x_n}$) into $k$
clusters ($C_1, ..., C_k$) in a way that
minimizes the within-cluster sum of squared distances.  Formally, the objective
function to minimize is
\begin{equation}
\sum_{i=1}^k \sum_{\boldsymbol{x} \in C_i} || \boldsymbol{x} - \boldsymbol{\mu_i} ||^2 ,
\end{equation}
where the mean or centroid of cluster $i$ is given by
\begin{equation}
\boldsymbol{\mu_i} = \frac{1}{|C_i|} \sum_{\boldsymbol{x} \in C_i} \boldsymbol{x},
\end{equation}
and $|C_i|$ is the size of cluster $i$.

\subsection{Silhouette analysis}
The user may specify the number of desired clusters arbitrarily or may
conduct a silhouette analysis to guide the choice of the number
desired clusters 1{\citep{rous87}}.  In a silhouette analysis,
one calculates the mean distance $a_j$ between point $j$ and all other
points in its cluster and the mean distance $b_j$ between point $j$
and all other points in the nearest cluster.  The silhouette value
\begin{equation*}
s_j = \frac{b_j - a_j}{\max{\{a_j,b_j\}}}
\end{equation*}
ranges between $-1$ and $+1$, where $-1$ indicates that point $j$ would be
more appropriately placed in the neighboring cluster, and $+1$ indicates
that point $j$ is appropriately clustered.  The mean silhouette value
over all data points is an indicator of the quality of the clustering
and can be used to determine the most appropriate number of clusters
to use for a data set.

\subsection{Clustering of Solar System bodies according to semimajor axis}
We explored k-means clustering of Solar System bodies to guide and
motivate a more general taxonomic system for planetary bodies.
{There is an important limitation to this approach. We used the
  Solar System because it is the only planetary system for which we
  have a sufficiently complete inventory to conduct this analysis.
  However, using a single system to derive a taxonomy applicable to
  many other systems is precarious.  It is entirely possible that the
  classification system will not generalize well and that revisions
  will be needed once additional information about exoplanetary
  systems becomes available.  In particular, it is possible that the
  Solar System is atypical, with perhaps a greater degree of order and
  stability than other systems, a condition that may be required for
  life to form and thrive.  With this caveat in mind, we
  investigated the taxonomic insights that can be garnered by
  clustering Solar System data.}

We started by evaluating the two features at our disposal when
classifying newly discovered planetary systems: orbital elements and
masses (Section~\ref{sec-practical}).  {To approximate the case
  of a newly discovered planetary system, we considered only the most
  massive bodies.}

\begin{figure}[h]
\begin{center}
  \includegraphics[width=\columnwidth]{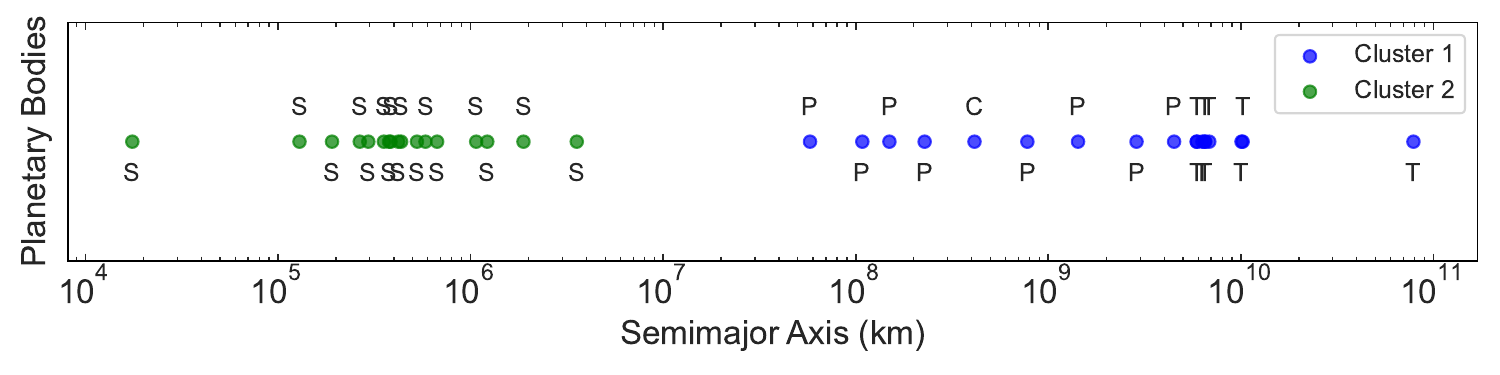}
\caption{K-means clustering of the semimajor axes of the 35 most
  massive planetary bodies in the Solar System.  The unsupervised
  clustering algorithm grouped satellites (S) in one cluster and
  all objects that orbit the Sun -- planets (P), Ceres (C), and trans-Neptunian primaries (T) -- in another.}
\label{fig-distance}
\end{center}
\end{figure}

We considered the 35 most massive planetary bodies in the Solar System
(8 planets, the Moon, Ceres, 4 Jovian satellites, 5 Saturnian
satellites, 5 Uranian satellites, Triton, and 10 trans-Neptunian
objects, including Pluto and Charon) and recorded the semimajor axes
of their orbits.  We applied k-means clustering to the logarithm (base
10) of the semimajor axes and conducted a silhouette analysis, which
indicated that the most suitable number of clusters in this data set
is two.  We verified that the optimal number of clusters and cluster
membership are robust against inclusion or exclusion of either one or
both of the most extreme objects (Charon and Sedna).  The unsupervised
clustering algorithm grouped all satellites in one cluster and all
objects that orbit the Sun in another (Figure \ref{fig-distance}).
{Other clustering algorithms (single linkage, Ward linkage, DBSCAN, Gaussian mixture model)} yielded the same result.

This finding motivates our first unsurprising conclusion with respect
to planetary taxonomy:
\begin{quote}
  {\bf Satellites are distinct from planets.}
\end{quote}
This taxonomic distinction is useful and desirable because it
emphasizes a relationship that is fundamental to the evolution of
planetary bodies.  Io's volcanic activity and Europa's distinctive
cycloids, for instance, would not exist if these worlds were in orbit
around the Sun.  Their orbit around Jupiter is an essential, defining
feature of their planetary identities.  We argue that such a
fundamental feature ought to be recognized in any planetary
taxonomy. Consequently, we strongly disfavor ``Moons are Planets''
taxonomies \citep[][]{metz22} because they obscure a fundamental
defining feature of planetary bodies.

{We are not asserting that planets and satellites will always be
  neatly divided into two clusters according to semimajor axis. For
  instance, we anticipate that the two groups may overlap when
  including short-period exoplanets, distant irregular satellites, or
  quasi satellites.  Likewise, the evidence for preferring the
  Gaussian mixture model with two clusters instead of four clusters is
  quite weak, with Bayesian information criterion (BIC) values of
  133.99 and 134.07, respectively.  Nevertheless, the planet-satellite
  distinction remains an important consideration when developing
  a planetary taxonomy.}

{We note that a distinction between satellites and planets is
  also expected when orbital distances to the primary are expressed in
  units of Hill radii or mutual Hill radii.  Most planets will orbit
  tens of (mutual) Hill radii away from the central body simply by
  virtue of the required spacing between neighboring planets.  Planets
  on orbits closer than $\sim$3.5 mutual Hill radii are not stable
  \citep{glad93,cham96}; the oligarchic growth stage during planet
  formation yields embryos spaced $\sim$10 mutual Hill radii apart
  \citep{koku02,thom03}; and the observed distribution of the spacings
  between neighboring exoplanets peaks near $\sim$20 mutual Hill radii
  \citep{fang13exopps, liss14}.  In contrast, satellites orbit planets
  at distances smaller than the Hill radius \citep{hami91}.}

\subsection{Clustering of Solar System bodies according to mass}

\begin{figure}[h!]
\begin{center}
  \includegraphics[width=\columnwidth]{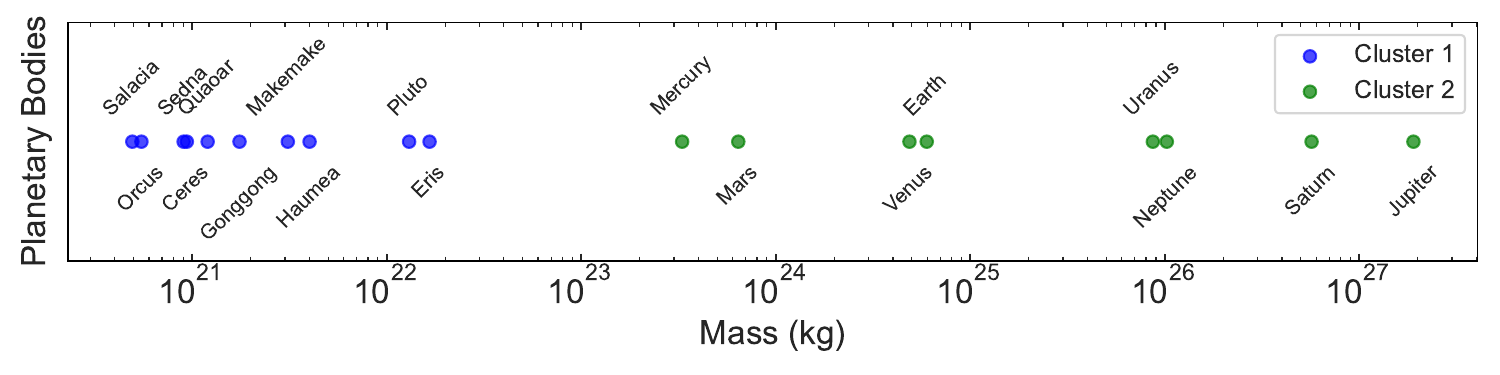}
\caption{Two-cluster k-means clustering of the masses of the 18 most massive
  planetary bodies that orbit the Sun.  The
  unsupervised clustering algorithm grouped the eight planets in one
  cluster and all remaining objects in another.}
\label{fig-2mass}
\includegraphics[width=\columnwidth]{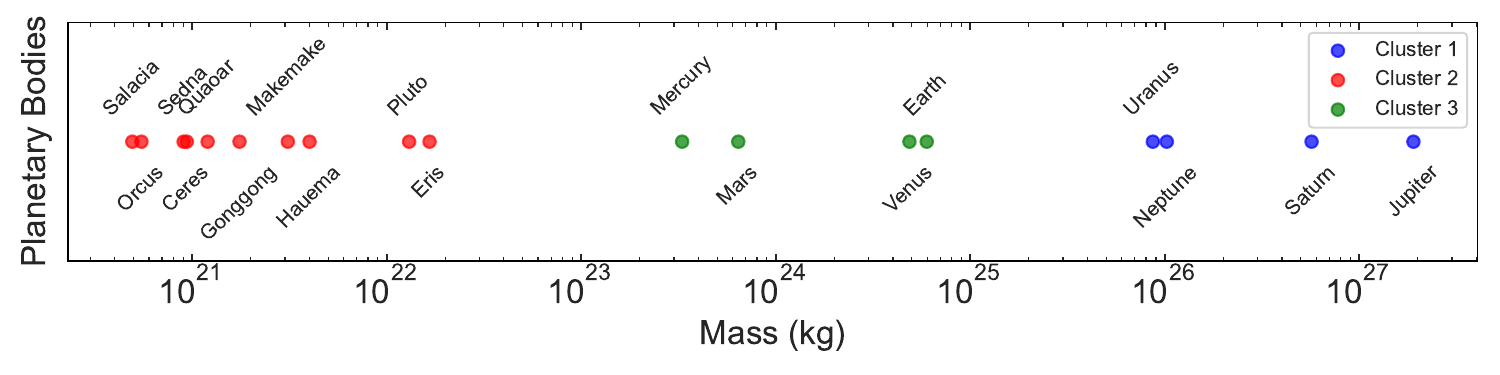}
\caption{Three-cluster k-means clustering of the masses of the 18 most massive
  planetary bodies that orbit the Sun.  The
  unsupervised clustering algorithm grouped the four giant planets in one cluster, the four terrestrial planets in another cluster, and all remaining objects in a third cluster.}
\label{fig-3mass}
\end{center}
\end{figure}

Having identified satellites as a distinct group, we continued our
exploration and applied k-means clustering to the logarithm (base 10)
of the masses of the 18 non-satellite objects (8 planets, Ceres, 9
trans-Neptunian objects, including Pluto).  The silhouette analysis
revealed that the data are best partitioned in two or three clusters,
with silhouette scores that are essentially tied (0.702 and 0.704,
respectively).  Figures~\ref{fig-2mass} and \ref{fig-3mass} reveal the
groupings, which are also unsurprising.  When asked to produce two
clusters, the unsupervised clustering algorithm grouped the eight
planets in one cluster and all remaining objects in another.  When
asked to produce three clusters, it grouped the four giant planets in
one cluster, the four terrestrial planets in another cluster, and all
remaining objects in a third cluster.  Another clustering algorithm
based on a Gaussian mixture model yielded the same results.  Either
grouping could be used to further develop a taxonomic system, with no
strong preference for either one.  However, as motivated by IAU resolution
B5, we refined our exploration to include a combination of orbital
elements and masses in order to diagnose dynamical dominance.

\begin{figure}[h]
\begin{center}
\includegraphics[width=\columnwidth]{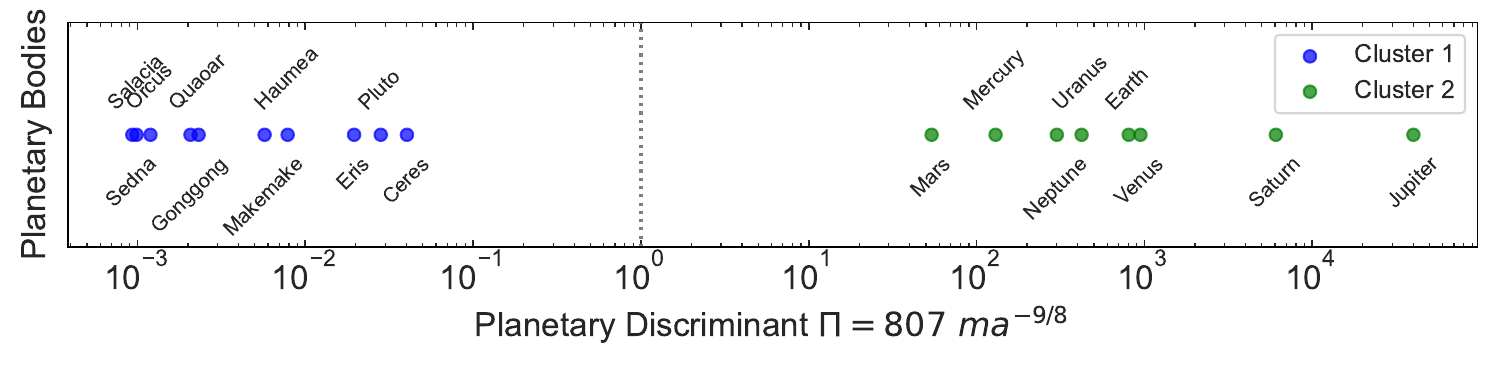}
\caption{K-means clustering of the planetary discriminant ($\Pi$) values of the 18 most massive
  planetary bodies that orbit the Sun.  The
  unsupervised clustering algorithm grouped the eight planets in one
  cluster and all remaining objects in another, with a three-order-of-magnitude gap between the groups.
The vertical dotted line represents the suggested threshold at $\Pi=1$.}

\label{fig-2Pi}
\end{center}
\end{figure}

\subsection{Clustering of Solar System bodies according to dynamical dominance}

The planetary discriminant $\Pi$
quantifies an object's ability to clear its feeding zone
in a given timescale
(Section~\ref{sec-Pi}), also known as dynamical dominance.  It can be
evaluated with knowledge of the mass of the planetary body and its
semimajor axis (Equation \ref{eq-Piss}).  Conveniently, both of these
quantities can generally be measured or estimated shortly after the
discovery of a planetary body.  We applied k-means clustering to the
logarithm (base 10) of the $\Pi$ values for the 18 non-satellite
objects.  The silhouette analysis showed that there are two clusters
in this data set.  
{For additional support of this conclusion, see Appendix A.}
Notably, all eight planets belong to the same group and
there is a gap of three orders of magnitude between the two groups
(Figure~\ref{fig-2Pi}).  A similar gap was previously identified and
used for taxonomic purposes by \citet{ster02} and \citet{sote06}.

This finding motivates our second conclusion with respect
to planetary taxonomy:
\begin{quote}
    {\bf Dynamical dominance provides a natural organizing principle for planetary taxonomy.}
\end{quote}
The taxonomic distinction provided by dynamical dominance is useful
and desirable because it identifies planetary bodies that
are
able to accrete or disperse most of the mass in their feeding zone, which
fundamentally governs their subsequent evolution.  For instance,
Earth's dynamical dominance allowed it to accrete 6$\times$10$^{24}$
kg of material, which ultimately enabled our planet to hold onto an
atmosphere and oceans, to differentiate, and to generate a magnetic
field.  Although these features are neither necessary nor sufficient
to label Earth a planet, they clearly distinguish Earth from other
planetary embryos near 1 au, which {did not accrete as much
  material as Earth and, as a result, were unable to experience}
a similar evolution.  {Collisional
  evolution also yielded different outcomes for Earth and the much
  less massive embryos nearby.}  Therefore, dynamical dominance is
another essential, defining feature of planetary identity.

We can be relatively confident that dynamical dominance is also a
general feature of planetary systems.  Planet formation simulations
consistently indicate that a small number of dynamically dominant
bodies emerge after a chaotic period of accretion
\citep{cham98,agno99,raym06}.

A dispassionate analysis of the features of Solar System bodies
yielded a distinct group of eight bodies that have been historically
referred to as planets.  Nothing in the clustering method is
engineered to exclude any particular object or to keep the number of
planets small.  The unsupervised clustering could have resulted in a
less numerous or more numerous group.

Readers who are chagrined that smaller bodies are not recognized as
planets should take comfort in the fact these these bodies
are no less worthy of exploration.  Indeed, some of the best
laboratories for studying planetary processes can be found on smaller
bodies.  In addition, our understanding of planetary systems would be
deficient without a careful study of planetary bodies of all sizes.
In other words, a taxonomic classification in one group or another is
not an indicator of scientific importance.

\section{A revised criterion for dynamical dominance}
\label{sec-simpler}

The unsupervised clustering of Solar System bodies suggests that a
reasonable approach to planetary taxonomy is to recognize satellites
and planets as distinct and to recognize planets as
capable of local dynamical dominance.
Dynamical dominance can be ascertained easily
because it is a simple function of mass and semimajor axis.

The choice of the clearing timescale in Equation (\ref{eq-mclear}) is
essentially the only arbitrary choice.
\citet{marg15planet} proposed to use the main-sequence lifetime of the
host star, which can be expressed as a simple function of stellar
mass.  Objects in orbit around brown dwarfs were explicitly excluded
from this initial formulation.  However, IAU Commission F2 included
brown dwarfs in its working definition, and we are interested in
developing a quantitative criterion that is fully consistent with this
working definition.  A timescale based on the main-sequence lifetime
of hydrogen-fusing stars must therefore be abandoned.

Here, we propose to adopt either the approximate main-sequence lifetime of
the Sun (10 billion years) or the current age of the universe (13.8
billion years) as a universal clearing timescale applicable to all
central bodies (brown dwarfs, stars, and stellar remnants).  
With these choices for the clearing timescale $t_{\rm clear}$, the orbit-clearing mass is
\begin{equation}
  m _{\rm clear} = 0.001239\, m_{\rm central}^{5/8}\, a^{9/8} \text{\ \ \ \ (for t$_{\rm clear}$ = 10 by),}
\label{eq-m10}
\end{equation}
or
\begin{equation}
  m _{\rm clear} = 0.000974\, m_{\rm central}^{5/8}\, a^{9/8} \text{\ \ \ \ (for t$_{\rm clear}$ = 13.8 by),}
\label{eq-m14}
\end{equation}
with units as in Equation (\ref{eq-mclear}).

The planetary discriminant is
\begin{equation}
\Pi = 807\, m\, m_{\rm central}^{-5/8}\, a^{-9/8} \text{\ \ \ \ (for t$_{\rm clear}$ = 10 by),} 
\label{eq-pi10}
\end{equation}
\begin{equation}
\Pi = 1027\, m\, m_{\rm central}^{-5/8}\, a^{-9/8} \text{\ \ \ \ (for t$_{\rm clear}$ = 13.8 by).} 
\label{eq-pi14}
\end{equation}
For t$_{\rm clear}$ = 10 by, values of $\Pi$ for the Solar System remain unchanged \citep[][Table 1]{marg15planet}.
For t$_{\rm clear}$ = 13.8 by, they are augmented by $\sim$27\%, a change that does not affect the classification of any Solar System body.

\begin{figure}[h!]
\begin{center}
\includegraphics[width=\columnwidth]{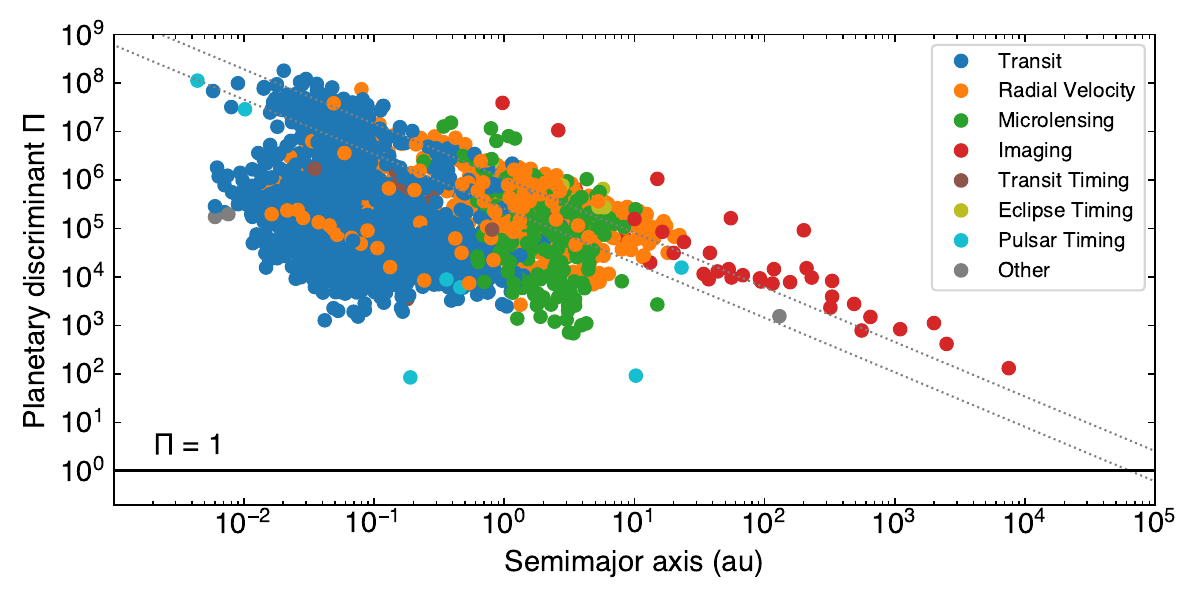}
\caption{Planet mass of confirmed exoplanets in units of the corresponding orbit-clearing mass for a clearing timescale t$_{\rm clear}$ = 10 by.  The solid black line shows the proposed boundary between planets and non-planets at $\Pi=1$.
  The dotted lines represent a Jupiter-mass planet in orbit around a 0.1 (top) and 1 (bottom) solar-mass star. Colors encode the discovery technique.  }
\label{fig-exo}
\end{center}
\end{figure}

We verified that all confirmed exoplanets known to date easily satisfy the criterion for dynamical dominance.  We downloaded the Planetary Systems Composite Data from the
\citet[][March 13 version]{nea13}, which contains information about 5595 confirmed exoplanets.  We eliminated objects with masses (or $M \sin{i}$) values larger than 13 Jupiter masses and computed values of $\Pi$ for the remaining 5443 objects (Figure~\ref{fig-exo}).
The smallest value is $\Pi \simeq 85$ for the pulsar planet PSR B1257$+$12 b with $a=0.190$ au and $m=0.02$ Earth masses.

\section{Proposed improvements to IAU Resolution B5 (2006)}
\label{sec-proposal}

Because it is relatively straightforward to apply a quantitative
orbit-clearing criterion to exoworlds, it is possible to extend the
2006 IAU planet definition to brown dwarfs and stars other than the
Sun and to remove ambiguity about what it means to
clear an orbital zone.  Likewise, it is possible to use a specific
mass threshold to replace a vague and impractical prescription regarding roundness.
We hope that these considerations will help start the conversation about making
planetary taxonomy both quantitative and useful.
One possible {IAU-aligned} formulation is as follows:
\enlargethispage{1.1cm}

\vspace{0.3cm}
{\em
A planet is a celestial body that
\begin{enumerate}[(a)]
\item orbits one or more stars, brown dwarfs, or stellar remnants, and
\item has sufficient mass to dynamically dominate the neighborhood around its orbit, i.e., 
$m > 0.0012\, m_{\rm central}^{5/8}\, a^{9/8}$,        
where $m$ is the mass of the planetary body expressed in Earth masses, $m_{\rm central}$ is the mass of the central body expressed is solar masses, and $a$ is the semimajor axis expressed in astronomical units, and
\item has sufficient mass for its self-gravity to overcome rigid body forces so that it is approximately in hydrostatic equilibrium and assumes a nearly triaxial shape, i.e., $m > 10^{21}$ kg, and
\item has a true mass below the limiting mass for thermonuclear fusion of deuterium (currently calculated to be 13 Jupiter masses for objects of solar metallicity), and
\item has a mass ratio with the central object below the L4/L5 instability, i.e.,  
$m/m_{\rm central} < 2/(25+\sqrt{621}) \simeq 1/25$.
\end{enumerate}

A satellite is a celestial body that orbits a planet.
}

\begin{figure}[h!]
\begin{center}
\includegraphics[width=0.53\columnwidth]{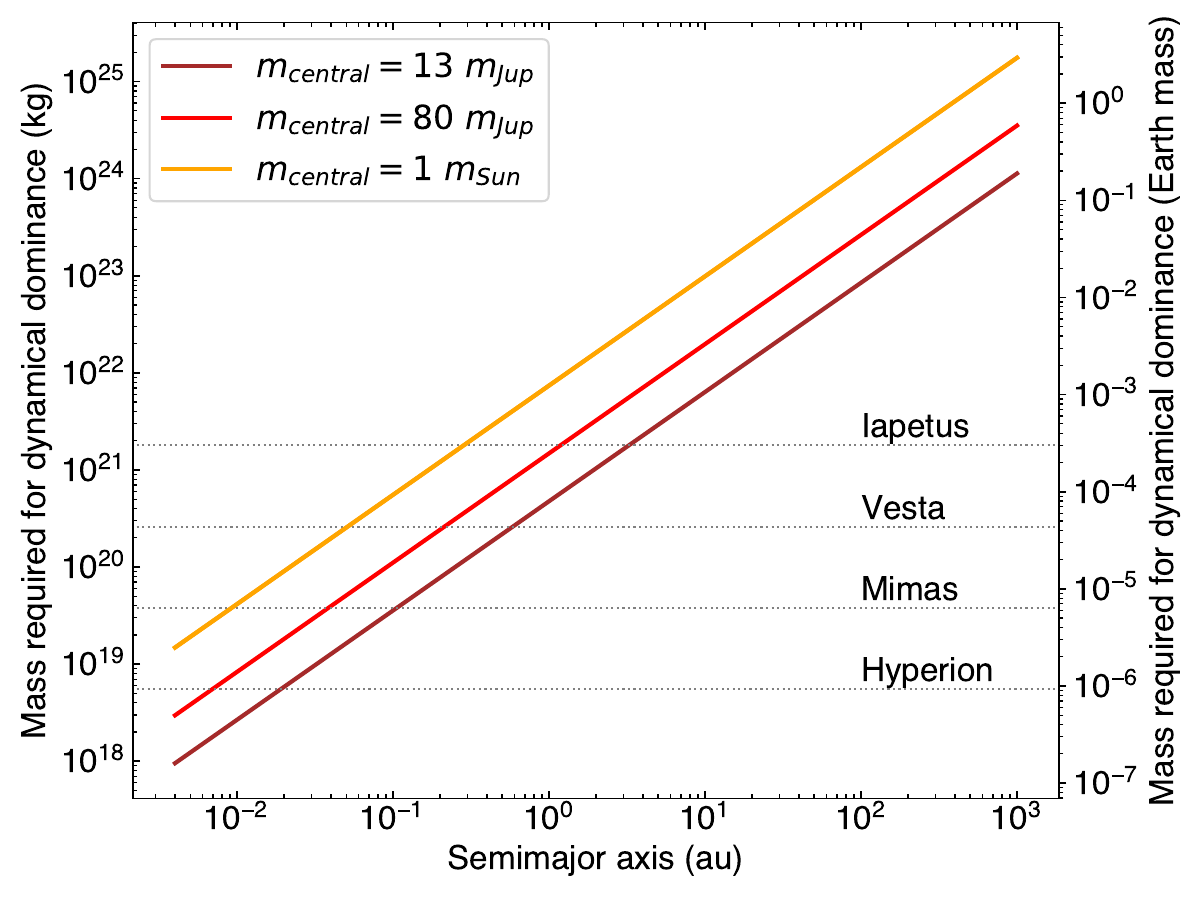}
\caption{Orbit-clearing mass as a function of semimajor axis for select values of the mass of the central body.  The masses of Iapetus, Vesta, Mimas, and Hyperion are shown with dotted lines.}
\label{fig-he}
\end{center}
\end{figure}

\vspace{0.3cm} In this formulation, clauses (a), (d), and (e) follow
the recommendations from the working definition of IAU Commission F2
\citep{leca22}.  {Clause (c) is a remnant of IAU Resolution B5
  (2006) and ``overcoming rigid body forces'' is also part of the IAU
  working definition of an exoplanet.  However, over a wide range of
  conditions, bodies that satisfy the dynamical dominance criterion
  also satisfy the hydrostatic equilibrium criterion
  \citep{marg15planet}, so clause (c) may be superfluous.  There
  may be exceptions, especially for bodies in close-in orbits around
  brown dwarfs.  For instance, a Mimas-mass object ($3.8\times10^{19}$
  kg) around a 13 Jupiter-mass brown dwarf is dynamically dominant at
  0.1~au but may not be in hydrostatic equilibrium
  (Figure~\ref{fig-he}). We encourage community conversations about
  the value of clause (c) and whether this clause could be
  eliminated.}

Two of us (J.L.M. and B.G.) submitted a proposed IAU resolution
(Appendix B)
in accordance with IAU procedures to IAU Commission F2 (Dec.\ 22, 2023 and Feb.\ 13, 2024) and IAU Division F (Jan.\ 17, 2024 and Feb.\ 13, 2024) with the goal of clarifying the definition of planet and extending it to exoplanets.
As of
{March 2024}, it appeared that the IAU had chosen not to publish the proposed resolution to allow for community feedback and voting at its 2024 General Assembly.
{Meaningful action} does not appear likely until the next General Assembly in 2027 at the earliest.

{
  \section{A Simpler Proposal}
  \label{sec-simplified}
  We recognize potential limitations and difficulties related to 
  clauses (b), (c), and (e) of the framework described in Section~\ref{sec-proposal}.}

{Although dynamical dominance is the cleanest criterion we have
  identified to classify planetary bodies (Figure \ref{fig-2Pi}), the
  semimajor axis dependence of the clearing mass (clause (b)) leads to
  the unfortunate consequence that the planetary status of a body
  depends on its distance from the central object.  Some of our
  colleagues regard this characteristic as problematic, although there
  is ample precedent for assigning location-dependent taxonomic labels
  to celestial objects (e.g., trans-Neptunian object vs.\ centaur
  vs.\ Jupiter-family comet, Amor vs.\ Apollo vs.\ Aten asteroid, and
  planet vs.\ free-floating planetary-mass object).}

{The shape-based aspects of the framework (clause (c)) are generally
  recognized as problematic.  For almost all practical purposes in the
  foreseeable future, we will not be able to obtain sufficient
  information about the shape of distant objects, and the criterion,
  if upheld, will likely reduce to a mass threshold.  However, a mass
  estimate is a poor predictor of an object's shape (Figure
  ~\ref{fig-round}).}

{The current framework includes a mass ratio (clause (e)) proposed by
  the IAU Working Group on Extrasolar Planets.  This ratio is also
  regarded as problematic because it would exclude Jovian planets
  around certain hosts from the planet category, e.g., a $>$4
  Jupiter-mass exoplanet orbiting an M9V star.  The necessity of this
  ratio, which was apparently recommended to distinguish objects with
  different formation mechanisms, is debatable.  Our planetary
  taxonomy might be more robust if we agreed to deemphasize formation
  hypotheses and instead agreed that a celestial object orbiting a
  brown dwarf is a planet, whether its mass is 1 Jupiter mass or 12
  Jupiter masses.}

{As a result of these three legitimate concerns, we also propose a
  much leaner definition with simple mass limits.  With such a
  definition, the connection to physical principles is less apparent,
  although we suggest a minimum mass that recognizes the gap between
  planets and non-planets in the Solar System (Figure~\ref{fig-2mass})
  and a maximum mass that recognizes the nominal deuterium burning
  limit.}
  
\vspace{0.3cm}
{\em
A planet is a celestial body that
\begin{enumerate}[(a)]
\item orbits one or more stars, brown dwarfs, or stellar remnants, and
\item is more massive than $10^{23}$ kg, and 
\item is less massive than 13 Jupiter masses ($2.5 \times 10^{28}$ kg).
\end{enumerate}}

{{\em A satellite is a celestial body that orbits a planet.}}
\vspace{0.3cm}

The IAU WG on Exoplanetary System Nomenclature has stated that freely
floating bodies (``rogue planets'') are not planets \citep{leca22}.  We
propose that rogue planets -- whatever they are called -- ought to
satisfy criteria (b) and (c).

\section{Conclusions}
{Precise definitions are needed to communicate and organize
  thoughts. The IAU definition of ``planet'' has been criticized with
  good reason since 2006.  Our community and the public deserve better
  definitions for such important astrophysical terms as ``planet'' and
  ``satellite''.}

{In this work, we examined the results of unsupervised clustering of
  Solar System bodies to propose guiding principles for planetary
  taxonomy.  This analysis revealed the presence of groupings among
  solar system bodies.  These groupings are not accidental and instead
  reveal profound differences in formation and evolution
  circumstances.  We suggest that our community would be better served
  with a quantitative taxonomic framework that recognizes such
  groupings and that is applicable to both Solar System bodies and
  exoworlds.  To facilitate further reflection on these topics, we
  have provided an IAU-aligned proposal and a simpler proposal.  These
  proposals are meant as a starting point for community conversations,
  and we welcome feedback on all aspects of both proposals.
}

\section{Acknowledgments}
{We thank Jack Lissauer and Eiichiro Kokubo for insightful suggestions.
We thank two reviewers, John Chambers and Jason W. Barnes, and the AAS Statistics Editor for
judicious comments that improved the manuscript.}

\appendix
\label{appendix}

{
\section{Additional support for two-cluster solution}
We applied a variety of unsupervised, nonparametric clustering
algorithms to the logarithm (base 10) of the $\Pi$ values for the 18
non-satellite objects: k-means, single linkage, Ward
linkage, and DBSCAN (eps=2).  All algorithms yielded the same
solution, suggesting that the two-cluster solution for $\Pi$ values is
robust.}

{Furthermore, we applied a Gaussian mixture model, which is a
parametric clustering algorithm, to the same data and computed the
Bayesian information criterion (BIC) values to assist in model
selection.  We found BIC values of 77.2, 84.3, 87.8, and 94.2 for 2,
3, 4, and 5 clusters, respectively.  The difference in BIC between the
two-cluster and three-cluster solutions is 7.1, which is considered
``strong evidence'' to prefer the former over the latter
\citep{kass95}.  In this case, the posterior odds are greater than 35
to 1.}

\section{Proposed resolution}

The text of the proposed resolution submitted 
{to IAU Commission F2 (Dec.\ 22, 2023 and Feb.\ 13, 2024) and IAU Division F (Jan.\ 17, 2024 and Feb.\ 13, 2024)}
is reproduced
below, {in part because it includes arguments in favor or revising the current definition.}  There are minor differences in notation with
{Section 6} of this
manuscript {(Mar.\ 27, 2024)} but the fundamental concepts are the same.
{Our manuscript also contains a simplified proposal in Section \ref{sec-simplified}.}

\vspace{0.3cm}
{\large \underline{Proposed Follow-up to IAU Resolution B5 (2006) ``Definition of a Planet in the Solar System''}}
\vspace{0.3cm}

\indent Proposers: Jean-Luc Margot (UCLA) and Brett Gladman (UBC)
\vspace{0.3cm}

{\bf \underline{Rationale}}
\vspace{0.3cm}

The current IAU definition of planet is inadequate due to vague (non-quantified) terms and it does not include exoplanets in a cohesive framework. The purpose of this document is to propose a clarification of the IAU definition of “planet” and an extension to exoplanets. A reasonable approach to astronomical nomenclature is to use physical principles to establish a meaningful threshold then adhere to the threshold for definitional purposes, as in the case of near-Earth asteroid dynamical classes. We propose both quantitative and non-quantitative versions of the resolution. In the quantitative version, clauses (d) and (e) are reproduced exactly from the Working Definition of the IAU WG on Exoplanetary System Nomenclature.

\vspace{0.3cm}
{\bf \underline{Resolution A (Quantitative)}}
\vspace{0.3cm}

The IAU resolves that a planet is a celestial body that\\
(a) orbits one or more stars, brown dwarfs, or stellar remnants, and\\
(b) has sufficient mass for its self-gravity to overcome rigid body forces so that it is approximately in hydrostatic equilibrium and assumes a nearly triaxial shape, and\\ 
(c) has sufficient mass to dynamically dominate the neighborhood around its orbit, i.e.,\\ 
\indent $M_p > 0.0012\, M_\star^{5/8}\, a_p^{9/8},$\\        
where $M_p$ is the mass of the planetary body expressed in Earth masses, $M_\star$ is the mass of the central body expressed is solar masses, and $a_p$ is the semi-major axis expressed in astronomical units, and\\
(d) has a true mass below the limiting mass for thermonuclear fusion of deuterium (currently calculated to be 13 Jupiter masses for objects of solar metallicity), and\\
(e) has a mass ratio with the central object below the L4/L5 instability, i.e.,\\  
\indent $M_p/M_\star < 2/(25+(621)^{1/2}) \simeq 1/25.$

\vspace{0.3cm}
{\bf \underline{Resolution B (Non-quantitative)}}
\vspace{0.3cm}

The IAU resolves that a planet is a celestial body that\\
(a) orbits one or more stars, brown dwarfs, or stellar remnants, and\\
(b) has sufficient mass for its self-gravity to overcome rigid body forces so that it is approximately in hydrostatic equilibrium and assumes a nearly triaxial shape, and\\
(c) has sufficient mass to dynamically dominate the neighborhood around its orbit, and\\
(d) has a true mass below the limiting mass for thermonuclear fusion of deuterium (currently calculated to be 13 Jupiter masses for objects of solar metallicity), and\\
(e) has a mass ratio with the central object below the L4/L5 instability (approximately 1/25). 

\vspace{0.3cm}
{\bf \underline{The 2024 GA Opportunity}}
\vspace{0.3cm}       

IAU Resolution B5 is problematic and the problems will not go away on their own. We have had 18 years to identify the problems and consider possible ways forward. There are good reasons to believe that we are better equipped in 2024 than in 2006 to produce a good outcome.  

\vspace{0.3cm}
(1) Increased transparency and community participation 
\vspace{0.3cm}

The 2006 IAU Planet Definition Committee did not allow for community review of its work and did not communicate its proposed planet definition to IAU members until after the start of the 2006 GA (via an embargoed press release released first to reporters). The proposal was flawed and had to be revised in relative haste during the short two-week span of the 2006 GA. This was ultimately the cause of the bad publicity and ensuing trauma. We have an opportunity to publicize a proposal prior to the GA and consider community feedback in a transparent and measured manner, without rushing the review process during the short time span of a GA. Both the IAU Resolutions Committee and the IAU Executive Committee have the prerogative to recommend approval or rejection of the proposed resolution (IAU Working Rules IV.18).

\vspace{0.3cm}
(2) Better voting representation
\vspace{0.3cm}

IAU procedures in 2006 required in-person voting and only a small fraction of IAU members (approximately 400 out of 9000 at the time) had the opportunity to vote on IAU Resolution B5. The IAU has since implemented an electronic voting system that allows broad participation and voting by IAU members.

\vspace{0.3cm}
(3) Improved clarity
\vspace{0.3cm}

IAU Resolution B5 is not quantitative and has generated profound disagreements about its interpretation. For instance, prominent planetary scientists have stated that Mercury and Venus would not be planets under IAU Resolution B5 because they may be slightly out of hydrostatic equilibrium, or that Earth and Jupiter may not be planets in that they have not have “cleared’’ the neighborhood around their orbits (because the term “cleared” is not quantitatively defined).  We have an opportunity to clarify the language of the resolution, address these criticisms, and possibly quantify what we mean.

\vspace{0.3cm}
(4) Link to exoplanets
\vspace{0.3cm}

The IAU Working Group on Exoplanetary System Nomenclature has promulgated a Working Definition of an Exoplanet that specifically refers to the Solar System: “The minimum mass/size required for an extrasolar object to be considered a planet should be the same as that used in our Solar System.” However, IAU Resolution B5 does not define what the minimum mass is. We have an opportunity to establish a unified definition for planets and exoplanets.   

\vspace{0.3cm}
(5) Elimination of an impossible standard
\vspace{0.3cm}

IAU Resolution B5 requires that a body has “cleared an orbit”. Complete clearing is an impossible standard to meet and a planet could lose its planetary status if a new small body began to cross the planet’s orbit. Evaluation of this criterion also requires difficult-to-attain knowledge about the existence or properties of other bodies in the vicinity (requiring a complete inventory of the entire system down to very small sizes). We have an opportunity to frame a definition that is aligned with IAU resolution B5’s orbit-clearing idea and to eliminate this impossible standard. We are proposing to use the concept of “dynamical dominance”, which is a natural organizing principle for classifying planets that relies on the ability to clear a zone and does not require instantaneous or complete clearing. A young body that is dynamically dominant may not yet have cleared its orbit but would still qualify as a planet because it is expected to clear its orbit over time (e.g., our Jupiter in the early planetesimal-clearing epoch).

\vspace{0.3cm}
(6) Congruence with the observed dichotomy of Solar System objects  
\vspace{0.3cm}

In the Solar System, planets and dwarf planets are cleanly separated by orders of magnitude 
by dynamical dominance criteria (e.g., Stern and Levison 2002, Soter 2006, Margot 2015). The figure below illustrates the latter criterion, which is used in clause (c).   

  \begin{center}
\includegraphics[width=0.6\columnwidth]{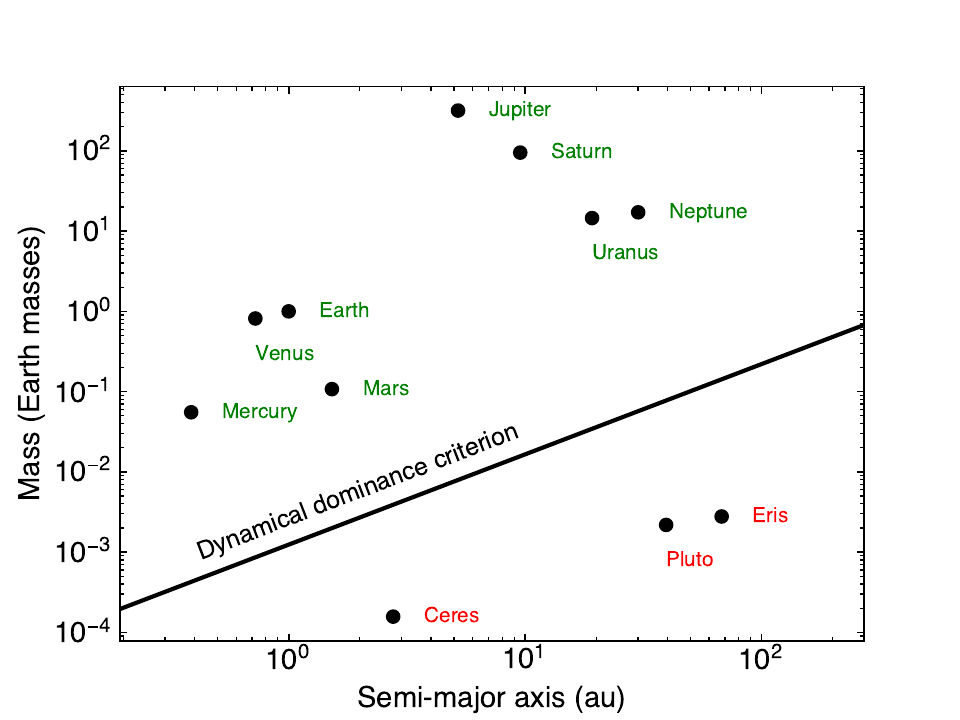}
\end{center}

We note that something like a hypothetical ``Planet Nine'' (e.g., $6.2^{+2.2}_{-1.3}$ Earth masses at $380^{+140}_{-80}$ au, per Brown and Batygin 2021) would satisfy the criterion for dynamical dominance.  

\vspace{0.3cm}
(7) Consistency with exoplanets
\vspace{0.3cm}

All known exoplanets with known estimates of mass and semi-major axis satisfy the dynamical dominance criterion.  

\vspace{0.3cm}
(8) Simplified classification process
\vspace{0.3cm}

The measurements required to establish dynamical dominance (stellar mass, planetary body mass, orbital period) are relatively easy to obtain and generally undisputed. In contrast, the measurements required to establish hydrostatic equilibrium (gravitational coefficients) are generally difficult to obtain without extensive radio tracking of an orbiting platform. Fortunately, in a very large fraction of the parameter space, the planetary bodies that have acquired sufficient mass to dynamically dominate their neighborhood are expected to also exceed the threshold for hydrostatic equilibrium (e.g., Margot, A Quantitative Criterion for Defining Planets, Astronomical Journal 150, 2015). Therefore, with an updated definition, we could make a determination about the planetary status of most exoplanets shortly after discovery (the exceptions are bodies orbiting low-mass stars or brown dwarfs at small semi-major axes).

\vspace{0.3cm}
(9) Opportunity for quantification
\vspace{0.3cm}

Part of the problems with the 2006 definition arose because of the well-meaning intent to produce a definition that the general public could easily understand. Our community may be better served with a quantitative definition. A quantitative approach requires that we ask the public to trust that professional astronomers have derived physically based, meaningful thresholds, even if the derivations of these thresholds may not be immediately obvious to non-astronomers. 

\vspace{0.3cm}
{\bf \underline{Potential limitations}}
\vspace{0.3cm}

Perfection is not attainable in the context of planetary nomenclature. We argue that physically motivated thresholds, even if imperfect, are superior to definitional voids or ambiguities.

\vspace{0.3cm}

-Circular orbits: Both the high-mass limit proposed by the IAU WG on Exoplanetary System Nomenclature and the low-mass limit proposed here rely on metrics derived for circular orbits. 
\vspace{0.3cm}

-Dynamical dominance criterion: A few criteria have been published to quantify dynamical dominance (e.g., Stern and Levison 2002, Soter 2006, Margot 2015). Our resolution is based on the latter, which involves clearing of the material in the canonical feeding zone of extent $2\sqrt{3}$ times the Hill radius, as opposed to ejection of the material to infinity.  

\vspace{0.3cm}

-Timescale for clearing: The proposed characteristic timescale for the ability to clear a zone in the dynamical dominance criterion is set to 10 billion years, which is comparable to the main-sequence lifetime of a solar-mass star. There is no profound meaning attached to this timescale and it could be equally set to, say, the current age of the universe. The coefficient in clause (c) comes from equation (8) in Margot, \href{http://doi.org/10.1088/0004-6256/150/6/185}{A Quantitative Criterion for Defining Planets}, Astronomical Journal 150, 2015 with $C=2\sqrt{3}$  and $t_\star$=10 by. With $t_\star$=13.7 by, the coefficient would be 0.0010.      

\vspace{0.3cm}

-Mass ratio: The mass ratio of 1/25 proposed by the IAU WG on Exoplanetary System Nomenclature may exclude certain planets (e.g., a $>$4 Jupiter-mass exoplanet orbiting a M9V star). The motivation for this threshold is explained in detail in Lecavelier des Etangs and Lissauer, \href{http://doi.org/10.1016/j.newar.2022.101641}{The IAU working definition of an exoplanet}, New Astronomy Reviews 94, 2022.

\vspace{0.3cm}

-Complexity of derivations: The derivation of the mass required to clear the canonical feeding zone and the derivation of the stability conditions for Lagrangian points are both within reach of professional astronomers.

\vspace{0.3cm}

-Inability to clear at large distances from the central body: With the proposed criterion, an Earth-mass body around a solar mass star is dynamically dominant out to 383 au. When orbital periods are long and the volume is enormous (at large distances), even an Earth-mass body does not dominate its neighborhood.

\vspace{0.3cm}

-In the quantitative proposal, clause (b) is currently non-quantitative. One could in principle use physical principles to establish a meaningful mass threshold for clause (b) as well.

\clearpage
\bibliography{bib}
\end{document}